\documentstyle[prl,aps,twocolumn,epsfig]{revtex}

\begin{document}
\title{Antiferromagnetic ordering in superconducting ${\bf YBa_2Cu_3O_{6.5}}$ }
\author{Y. Sidis$^{1}$, C. Ulrich$^{2}$, P. Bourges$^{1}$, C. Bernhard$^{2}$, C.
Niedermayer$^{3}$, L.P. Regnault$^{4}$, N.H. Andersen$^{5}$ and B.~Keimer$%
^{2}$}
\address{1 - Laboratoire L\'{e}on Brillouin, CEA-CNRS, CE Saclay, 91191 Gif
sur Yvette, France}
\address{2 - Max-Planck-Institut f\"{u}r Festk\"{o}rperforschung, 70569 Stuttgart,
Germany }
\address{3 - University of Konstanz, 78434 Konstanz, Germany }
\address{4 - CEA Grenoble, D\'{e}partement de Recherche Fondamentale sur la
mati\`{e}re Condens\'{e}e, 38054 Grenoble cedex 9, France}
\address{5 - Condensed Matter Physics and Chemistry Department, Ris\/o National
Laboratory, DK-4000 Roskilde, Denmark}
\draft
\maketitle

\begin{abstract}
Commensurate antiferromagnetic ordering has been observed in the
superconducting high-$Tc$ cuprate ${\rm YBa_{2}Cu_{3}O_{6.5}}$ (${\rm T_{c}}$%
=55 K) by polarized and unpolarized elastic neutron scattering. The magnetic
peak intensity exhibits a marked enhancement at $T_{c}$. Zero-field $\mu $SR
experiments demonstrate that the staggered magnetization is not truly static
but fluctuates on a nanosecond time scale. These results point towards an
unusual spin density wave state coexisting with superconductivity.
\end{abstract}


\vskip .5 cm 



The coexistence of superconductivity with an antiferromagnetic (AF) state
has recently been reported in certain Ce based heavy fermion systems under
pressure \cite{Mature98}, inspiring theories of spin fluctuation mediated
pairing of electrons in these systems. In contrast, the antiferromagnetic
and superconducting phases of the copper oxide high-$T_{c}$ superconductors
are generally considered to be well separated. However,
since the early days of high temperature superconductivity, there have been
persistent reports of local magnetic moments in the metallic and
superconducting regimes of the phase diagram. Specifically, zero-field muon
spin resonance (ZF-$\mu $SR) measurements \cite{Kiefl89,Niedermayer97} in
both ${\rm La_{2-x}Sr_{x}CuO_{4}}$ and ${\rm Y_{1-x}Ca_{x}Ba_{2}Cu_{3}O_{6}}$
compounds provide indications of a microscopic coexistence of
superconductivity ($T_{c}\leq $ 50 K) and frozen magnetic moments at low
temperature ($T\leq $ 10 K), and M\"{o}ssbauer spectroscopy shows low
temperature spin freezing up to x $\leq $ 0.66 in ${\rm YBa_{2}Cu_{3}O_{6+x}}
$ \cite{Hodges}. Both $\mu $SR and M\"{o}ssbauer spectroscopies are local
probes that provide no information about the spatial correlations of the
magnetic moments. An early observation of coexistence between
superconductivity and antiferromagnetism from neutron measurements in ${\rm %
YBa_{2}Cu_{3}O_{6.55}}$ has remained unconfirmed \cite{Petitgrand}. More
recently, an incommensurate magnetic ordering below a critical temperature $%
{\rm T_{m}}$ has been observed by elastic neutron scattering in ${\rm %
La_{2}CuO_{4}}$-family superconductors \cite{Lee99}. Interestingly, this
magnetic phase coexists with superconductivity in a certain doping range. In
super-oxygenated ${\rm LaCuO_{4+\delta }}$, ${\rm T_{m}}$ and ${\rm T_{c}}$
are actually identical (${\rm T_{c}}$=42 K), whereas ${\rm T_{m}}$ remains
larger than ${\rm T_{c}}$ in ${\rm La_{1.6-x}Nd_{0.4}Sr_{x}CuO_{4+\delta }}$
compounds for x$\leq $ 0.2.

Here we report the observation, by neutron scattering, of {\it commensurate}
AF order in the superconducting compound ${\rm YBa_{2}Cu_{3}O_{6.5}}$ ($T_{c}
$=55 K), using polarized and unpolarized elastic neutron scattering. This AF
order develops at a high temperature (${\rm T_{N}}$=310 K) and is
characterized by a large correlation length around 100\AA , but the low
temperature staggered magnetization remains weak, ${\rm m_{0}=0.05\mu _{B}}$%
. Observation of such a small moment with a large ${\rm T_{N}}$ is
reminiscent of a SDW state observed in itinerant magnets \cite{Moriya}.
Interestingly, the magnetic signal around the AF Bragg reflections exhibits
a significant enhancement at ${\rm T_{c}}$. Finally, ZF-$\mu $SR
measurements carried out on the same sample show that the AF\ staggered
magnetization is not truly static but fluctuates with a characteristic
frequency in the $\mu $eV range.

The experiments were performed on a large ${\rm YBa_{2}Cu_{3}O_{6.5}}$
single crystal ($\sim $ 23 g) grown using the top seed melt texturing method
\cite{Fong99}. The sample was subsequently annealed to achieve an oxygen
concentration of $x=0.5$ \cite{O-Treatment}. We checked by neutron
measurements at room temperature the presence of the $Q=(2.5,0,5)$ Bragg
peak characteristic of the {\it orthorhombic}-II phase \cite{O-Treatment}
and derive the correlation lengths: $\xi _{ab}\simeq $ 20 \AA\ and $\xi
_{c}\simeq $ 12 \AA . This demonstrates well ordered CuO chains and hence
uniform hole doping within the ${\rm CuO_{2}}$ planes. The sample displays a
sharp superconducting transition at ${\rm T_{c}}$=55 K, measured by a
bulk-sensitive neutron depolarization technique \cite{Fong99}. Neutron
scattering experiments were carried out on the 4F1 and 4F2 cold triple axis
spectrometers at the reactor Orph\'{e}e in Saclay (France). These
spectrometers are equipped with double monochromators consisting of two
Pyrolithic Graphite (PG) crystals whose (002) reflections were set to
select the incident wave vector $k_{I}$= 1.48 \AA $^{-1}$. For unpolarized
neutron scattering measurements we used a PG(002) analyzer. Two filters, Be
and PG, were used to remove higher order contamination. Polarized neutron
scattering experiments were performed with a bender to polarize the incident
beam and with a Heusler (111) analyzer. A conventional polarized neutron
setup \cite{Moon} was used with a flipper in the scattered beam to rotate
the neutron spin polarization by 180$^{\circ }$ and a magnetic guide field
{\bf H} to preserve the neutron polarization. The (110) and (001) directions
of reciprocal space were within the horizontal scattering plane. We quote
the scattering vector ${\bf Q}=(H,K,L)$ in units of the reciprocal lattice
vectors {\bf a}$^{\ast }\sim $ {\bf b}$^{\ast }$=1.63 \AA $^{-1}$ and {\bf c}%
$^{\ast }$=0.53 \AA $^{-1}$.

\begin{figure}[h]
\epsfxsize=7.5 cm $$ \epsfbox{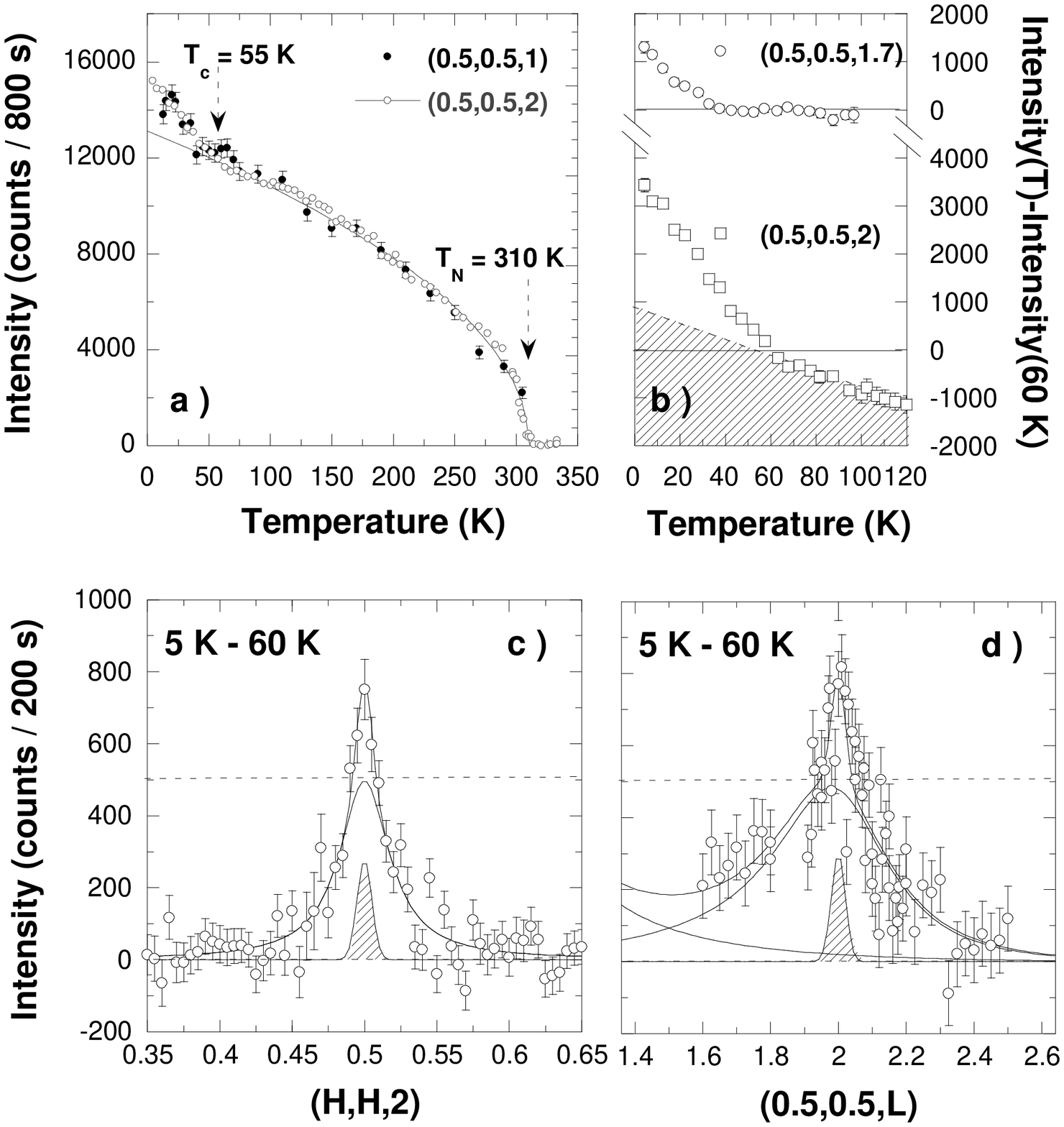} $$
\caption{ a) Temperature dependence of the magnetic intensity measured at $%
{\bf Q}=(0.5,0.5,2)$ (with an unpolarized beam) and $(0.5,0.5,1)$ (with a
polarized beam) scaled to each other. From ${\rm T_{N}} $=310 K down to $%
{\rm T_{c}}$=55 K, the magnetic intensity increases as a power law (dashed
line). b) Enhancement of the low temperature static magnetic response at $%
{\bf Q}=(0.5,0.5,2)$ and $(0.5,0.5,1.7)$ with respect to the magnetic
scattering at 60 K. The dashed line and the shaded area extrapolate down to
low temperature the temperature dependence of the $(0.5,0.5,2)$ magnetic
Bragg reflection as measured above ${\rm T_{c}}$ (panel a). c and d)
Difference between scans at T=5 K and T=60 K measured around $(0.5,0.5,2)$
along (110) and (001), respectively. }
\label{Fig1}
\end{figure}

The elastic signal was discovered in the course of unpolarized elastic
neutron scattering experiments on superconducting ${\rm YBa_{2}Cu_{3}O_{6.5}}
$ around the wavevectors ${\bf Q}=(0.5,0.5,L)$ for $L\neq 0$ integer. The
structure factor of this diffraction pattern is identical to that of the
undoped AF parent compound ${\rm YBa_{2}Cu_{3}O_{6}}$. Fig.~\ref{Fig1}a
shows the temperature dependence of \ the intensity measured at two of these
wave vectors: $L$=1,2. The strong temperature dependence, as well as the
presence of two filters in the beam, ensures that the signal is not due to $%
\lambda /2$ contamination, and extensive polarized beam experiments (below)
demonstrate that the signal is indeed of magnetic origin. The intensity
exhibits a sharp onset at a ``N\'{e}el'' temperature of ${\rm T_{N}}$=310 K.
From ${\rm T_{N}}$ down to ${\rm T_{c}}$=55 K, the magnetic intensity
continuously increases, following the power law: $(1-T/T_{N})^{2\beta }$
with $\beta $=0.25 (dashed line in Fig.~\ref{Fig1}a). 
 Surprisingly, another marked upturn of
the AF\ intensity is observed at the {\it superconducting} transition
temperature ${\rm T_{c}}$ (Fig.~\ref{Fig1}b). After calibration against
nuclear Bragg reflections, the staggered moment at T=60 K is found to be $%
{\rm m_{0}\sim 0.05\mu _{B}}$ (assuming a homogeneous distribution of the
magnetic moments at all copper sites in the plane), that is, more than an
order of magnitude smaller than in the insulating parent compound ${\rm %
YBa_{2}Cu_{3}O_{6}}$ \cite{Rossat88,Shamoto}. The small moment (which
translates into an elastic magnetic cross section more than two orders of
magnitude smaller than that of ${\rm YBa_{2}Cu_{3}O_{6}}$) explains why the
signal could not be measured in previous neutron measurements on smaller
samples.

We next describe polarized beam experiments designed to confirm the magnetic
origin of the signal and to determine the moment direction. The matrix
element for magnetic neutron scattering can be written as \cite{Lovesey}
\begin{equation}
\langle m^{\prime }|\overrightarrow{\sigma }\bullet \overrightarrow{M}_{\bot
}|m\rangle
\end{equation}

where $|m\rangle ,|m^{\prime }\rangle =|\pm
{\frac12}%
\rangle $ are the initial and final states of the neutron spin, $%
\overrightarrow{\sigma }$ is the neutron spin Pauli matrix, and $%
\overrightarrow{M}_{\bot }$ is the component of the electronic magnetic
moment perpendicular to the scattering vector {\bf Q}. Note that the
functional form of the matrix element does not depend on whether $%
\overrightarrow{M}$ originates in the spin or the orbital motion of the
electrons. The direction of the neutron spin quantization axis at the sample
position is selected by a small guide field {\bf H. }From Eq. (1) one sees
that the magnetic intensity is entirely spin-flip (SF) when {\bf H}//{\bf Q}%
. When {\bf H}$\perp ${\bf Q}, on the other hand, the ratio of magnetic
intensities in SF\ and non-spin-flip (NSF) channels depends on the
orientation of $\overrightarrow{M}$.

\begin{figure}[h]
\epsfxsize=7.5 cm $$ \epsfbox{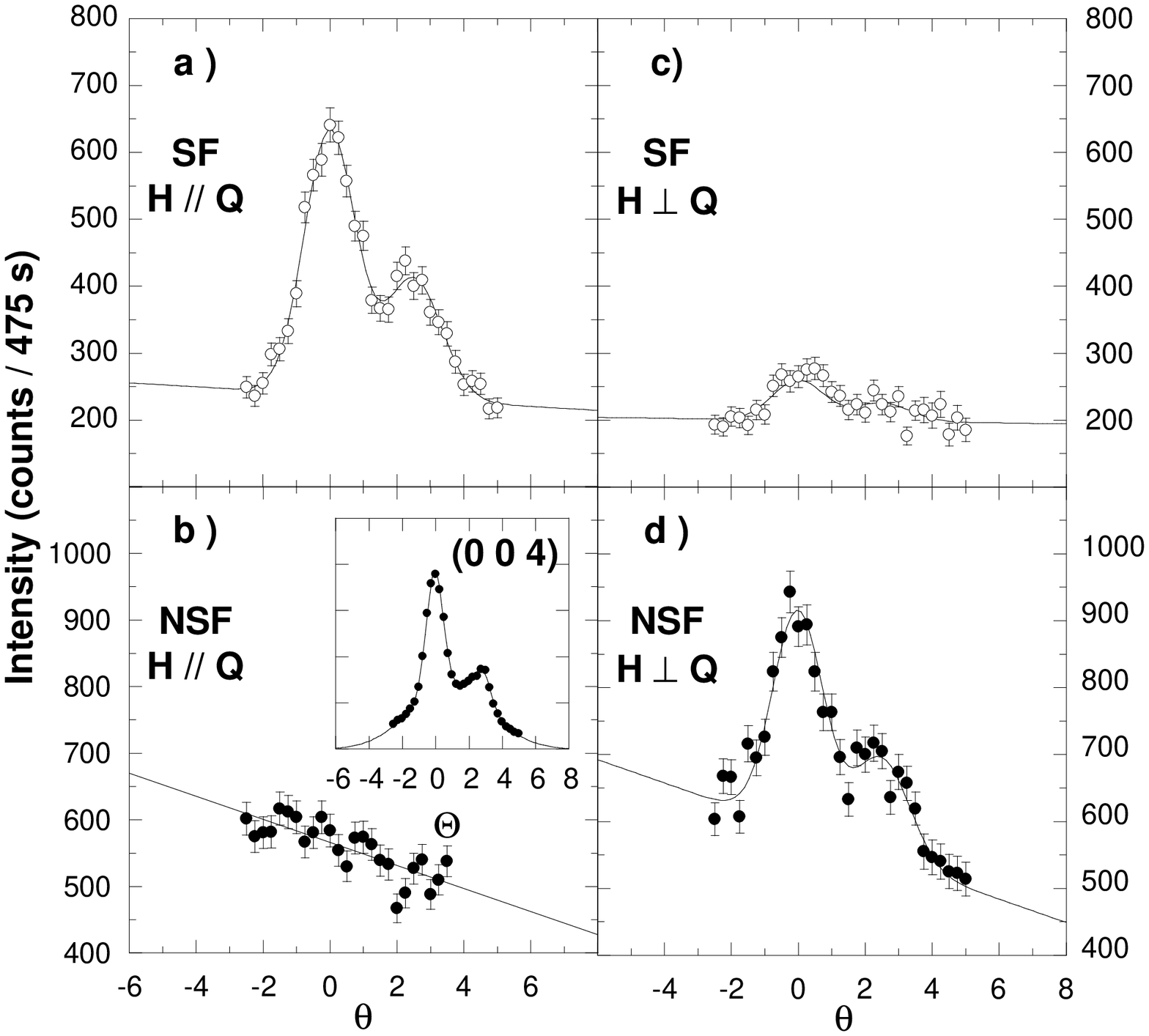} $$
\caption{ Polarized neutron scattering measurements at T=60 K, showing
rocking scans at ${\bf Q}=(0.5,0.5,1)$ in both SF and NSF channels for two
different magnetic guide field directions (see text). The inset shows a
rocking scan at the nuclear Bragg reflection (1,1,0), measured by
unpolarized neutron scattering. The double peak structure originates from
two distinct crystallites in the sample.}
\label{Fig2}
\end{figure}

Figs. \ref{Fig2}a-d show rocking scans performed around ${\bf Q}=(0.5,0.5,1)$
at T= 60 K. The two-peak profile is due to two crystallographic grains in
our sample (a rocking scan through a nuclear reflection with the same
profile is shown in the inset). For {\bf H}//{\bf Q} a peak is seen only in
the SF channel (Fig.~\ref{Fig2}a), while the measured scattering intensity
in the NSF channel remains featureless (Fig.\ref{Fig2}b). This proves that
the observed signal is of magnetic origin. The results for {\bf H}$\perp $%
{\bf Q} in Figs.\ref{Fig2}c-d then provide information about the orientation of
the magnetic moments. The full elastic magnetic cross section is
proportional to $\frac{1}{2}\langle M\rangle _{a,b}^{2}(1+\sin ^{2}\theta
_{l})+\langle M\rangle _{c}^{2}\cos ^{2}\theta _{l}$, where ${\it \theta }%
_{l}$ stands for the angle between {\bf Q} and the (110) direction ($\theta
_{l}$=25$^{\circ }$ for ${\bf Q}=(0.5,0.5,1)$), and $\langle M\rangle _{a,b}$
and $\langle M\rangle _{c}$ represent the thermodynamic averages of the
magnetization within the $a,b$ plane and perpendicular to it, respectively.
(Note that {\it a} and {\it b} directions are superposed in our experiment
due to twinning). For {\bf H} along (1$\overline{1}$0) perpendicular to the
scattering plane (Figs.\ref{Fig2}c-d), this intensity is apportioned such
that the SF channel measures $\langle M\rangle _{a,b}^{2}\frac{1}{2}\sin
^{2}\theta _{l}+\langle M\rangle _{c}^{2}\cos ^{2}\theta _{l}$ and the NSF
channel measures $\frac{1}{2}\langle M\rangle _{a,b}^{2}$. The weak
intensity observed in the SF channel (Fig.\ref{Fig2}c) and the larger
intensity in the NSF channel demonstrate that the magnetic moments are
predominantly within the basal $a,b$ plane. Additional measurements with
{\bf H}$\perp ${\bf Q} but for {\bf H} in the scattering plane corroborate
this conclusion.

Both the moment direction and the structure factor of the magnetic order
observed in ${\rm YBa_{2}Cu_{3}O_{6.5}}$ are thus similar to those of the
undoped parent compound. One possible interpretation of this observation,
namely macroscopic or mesoscopic concentration gradients of oxygen leading
to an inhomogeneous charge distribution, can be ruled out. First, the shapes
of the rocking curves around the AF Bragg reflections (Fig.~\ref{Fig2}a,c,d)
exactly reproduce those of the nuclear Bragg reflections (insert in Fig.~\ref
{Fig2}b). The observed double peak structure originates from two distinct
grains in our sample, which are separated by a tilt angle of 2.5$^{\circ }$.
The volume ratio of these grains is found the same for structural and
magnetic scattering, ruling out phase segregation where magnetic order only
occurs in a small part of the sample. Second, we observe a sharp transition
with $T_{N}=$ 310 K whereas in the case of an inhomogeneous sample one would
expect a broad distribution of transition temperatures starting from $T_{N}=$
410 K, the N\'{e}el temperature of the undoped compound. Conversely, the
superconducting transition of our sample measured by a bulk-sensitive
technique is also very sharp. Third, we observe a marked increase of the
magnetic intensity at $T_{c}$; a minority phase of the undoped compound
would not be affected by superconductivity. Finally, while the N\'{e}el
state in the undoped insulator is associated with a static staggered
magnetization, the $\mu $SR measurements described below show that such a
static moment is absent in ${\rm YBa_{2}Cu_{3}O_{6.5}}$. These observations
imply the absence of large-scale inhomogeneities in our sample.

A quantitative analysis of Figs.~\ref{Fig2}a and~\ref{Fig2}d reveals a
systematic broadening of the rocking curve around the AF Bragg reflection $%
{\bf Q}=(0.5,0.5,1)$ with respect to the nuclear one, indicating finite size
AF correlations. At 60 K, an anisotropic correlation length, $\xi _{a}\simeq
20a$ and $\xi _{c}\simeq 9c$, is determined from scans along (110) and (001)
around the AF Bragg reflections. From 60 K up to $T_{N}$, these correlation
lengths remain constant within the error bars. At 60 K, the intensity ratio
(peak amplitude) between the AF Bragg reflections I$(L=1)$/I$(L=2)=0.67\pm
0.05$ is somewhat larger than expected for localized (${\rm d_{x^{2}-y^{2}}}$%
) Cu spins aligned in the plane \cite{Shamoto,Rossat93}. The analysis of the
magnetic structure factor over several AF peaks (not presented here)
actually shows a form factor decreasing faster at large $|Q|$, consistent
with more delocalized magnetic states.

Below {\it T}$_{c}$, we observe a pronounced {\it broadening }of the
magnetic reflections that goes along with the upturn in the peak intensity.
We have performed scans around {\bf Q}=(0.5,0.5,2) along both the (110) and
the (001) directions at T= 5 K and T= 60 K whose differences are reported in
Fig.~\ref{Fig1}.c and Fig.~\ref{Fig1}.d, respectively. The additional low
temperature intensity remains centered at ${\bf Q}=(0.5,0.5,2)$, but because
of the broadening it becomes observable all along ${\bf Q}=(0.5,0.5,L)$, for
instance at ${\bf Q}=(0.5,0.5,1.7)$ as shown in Fig.~\ref{Fig1}b. The
observed momentum broadening cannot simply be attributed to a decrease of
the AF correlation length in the superconducting phase, because the peak
intensity at ${\bf Q}=(0.5,0.5,2)$ intensity shows an {\it upturn} below $%
T_{c}$, rather than the downturn expected if the integrated intensity
remained constant. It is therefore more appropriate to consider the
development of a second type of AF order (with much shorter correlation
length) below ${\rm T_{c}}$, in addition to the one already present in the
normal state. Figs.~\ref{Fig1}c-d have thus been fitted as a superposition
of two contributions centered at the wave vector ${\bf Q}=(0.5,0.5,2)$. The
sharper one, with a Gaussian profile with ${\rm \Delta q}$=0.0135 \AA ${\rm %
^{-1}}$, corresponds to the continuous increase of the magnetic response
observed above ${\rm T_{c}}$ (shaded peaks in Fig.~\ref{Fig1}c-d). The
second contribution, with a broader Lorentzian profile, is characterized by
an onset at ${\rm T_{c}}$ and an anisotropic correlation length: ${\rm \xi
_{ab}\simeq }$ 22 \AA\ (${\rm \Delta q}$=0.02 \AA $^{-1}$, Fig.~\ref{Fig1}%
.c) and ${\rm \xi _{c}\simeq }$ 9 \AA\ (${\rm \Delta q}$=0.046 \AA $^{-1}$,
Fig.~\ref{Fig1}d). It is noteworthy that the superconducting coherence
length, ${\rm \xi ^{SC},}$ in high-${\rm T_{c}}$ superconductors is
similarly anisotropic; in particular ${\rm \xi _{ab}^{SC}\sim }$ 20\AA\ and $%
{\rm \xi _{c}^{SC}\sim }$ 3-7 \AA \cite{SC} in ${\rm YBa_{2}Cu_{3}O_{6+x}}$.
The second order parameter below ${\rm T_{c}}$ accounts for only 15 \% of
the overall magnetic peak intensity (Fig.~\ref{Fig1}.a), but because of its
broader structure in q-space its integrated intensity (${\rm \sim 0.07\mu
_{B}}$) is comparable to that of the first component above ${\rm T_{c}}$.

\begin{figure}[t]
\epsfxsize= 7.5 cm $$\epsfbox{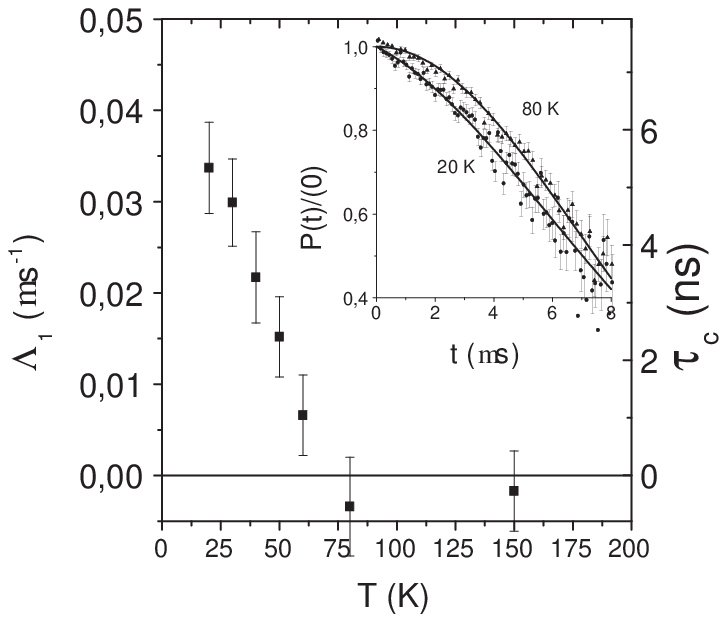}$$
\caption{ Temperature dependence of the ZF-$\protect\mu $SR relaxation rate $%
\Lambda _{1}$ (left scale) and relaxation rate of the magnetic moments $%
\protect\tau _{c}$ (right scale). Inset: time dependent muon spin
polarization at T=20 K and 80 K. }
\label{Fig3}
\end{figure}

Additional higher resolution (smaller $k_{I}$) measurements show that the
signal remains resolution limited in energy when the resolution is tightened
to $\sim $50 $\mu $eV. Thus, the observed AF order appears static on a time
scale shorter than $\sim $10$^{-10}$ s. In order to obtain sensitivity to
spin fluctuations on a larger time scale, ZF-$\mu $SR measurements were
performed at the GPS beamline of the Paul-Scherrer-Institute (PSI) in
Villigen, Switzerland, on a piece cut from the same sample. For T$>$60 K
only a very slow depolarization of the muon spin polarization $P(t)$ is
observed which is due to the nuclear Cu moments and is well described by a
Kubo-Gauss function (KG) with a damping rate of 0.12 $\mu s^{-1}$\cite
{Niedermayer97}. Static AF ordered moments of size 0.05 $\mu _{B}$ (as
indicated by the neutron data) should give rise to an oscillating signal
with a precession frequency of about 0.4 MHz, that is, about 1/10 the value
that is typically observed in ${\rm YBa_{2}Cu_{3}O_{6}}$ \cite{Niedermayer97}%
. Even under the assumption that these static electronic magnetic moments
are strongly disordered, we should still observe a rapid depolarization with
a rate of $\Lambda _{1}\sim 2.5{\rm \mu s}^{-1}$. The circumstance that this
is clearly not observed implies that the magnetic moments fluctuate on a
time scale longer than the one of the neutron scattering experiment ($%
10^{-10}$ s) but much shorter than the one of the $\mu $SR experiment ($%
10^{-6}$ s). As shown in the inset of Fig. \ref{Fig3}, the depolarization
becomes somewhat faster at low temperatures, and the shape of $P(t)$
gradually changes from Gaussian to exponential. The inset of Fig. \ref{Fig3}
shows the result of a fit using the function $P(t)=P(0)\times KG\times \exp
(-\Lambda _{1}t)$. The KG-function describes the contribution of the nuclear
Cu-moments which should be T-independent; the KG depolarization rate thus
has been fixed to the value at 150 K, i.e. 0.12 ${\rm \mu s^{-1}}$. The
exponential function describes the depolarization due to the rapidly
fluctuating electronic moments. Figure \ref{Fig3} shows the T-dependence of $%
\Lambda _{1}$ as given by the scale on the left hand side. For the limit of
rapidly fluctuating moments having a characteristic relaxation rate $\tau
_{c}$ ($\gamma _{\mu }B_{\mu }<<\tau _{c}$) and under the assumption that
the field $B_{\mu }\approx 40$ G at the muon site is determined by the
neutron scattering measurements above, we obtain $\tau _{c}=\Lambda
_{1}/(\gamma _{\mu }B_{\mu })^{2}=\Lambda _{1}/6.3$ [$\mu$s] as shown by the
scale on the right hand side of Fig. 3.

In summary, an unusual commensurate AF phase, with in-plane magnetic moments
fluctuating on a nanosecond time scale, is found to coexist with
superconductivity in ${\rm YBa_{2}Cu_{3}O_{6.5}}$. As in itinerant magnetic
systems \cite{Moriya}, we observe a small value of the ordered moment ($\sim
{\rm 0.05\mu _{B}}$ at T=60 K) together with a large $T_{N}$= 310 K. The
observed commensurate structure is not compatible with stripe ordering whose
structure in momentum space is incommensurate. A disorder-induced AF\
long-range ordered state with spatially inhomogeneous local moment has been
predicted in the spin gap phase of the cuprates \cite{fukuyama}. The
coexistence of an AF SDW state and d-wave superconductivity has also been
considered theoretically \cite{Murakani98,Bouis99,af}. It has been shown
that when both states coexist there are unusual coherence effects and a $\pi
$-triplet superconducting order parameter appears at ${\rm T_{c}}$ (when $%
{\rm T_{c}\leq T_{N}}$). Electrons in the superconducting condensate may
thus contribute the unusual neutron scattering response we observe below $%
{\rm T_{c}}$. Finally, it was shown very recently that the staggered orbital
current patterns proposed earlier \cite{orbital} can produce in-plane
magnetic moments under certain conditions \cite{chakravarty}.

We acknowledge stimulating discussions with S. Chakravarty, J. Hodges, G.
Khaliullin, F. Onufrieva and P. Pfeuty.



\end{document}